\documentclass[twocolumn,twoside,slac_two]{revtex4}
\usepackage{graphicx}
\usepackage{fancyhdr}
\begin{document}

\title{Recent Results in Bottomonium\\
Ties to Charmonium}

%

\author{D. Kreinick}
\affiliation{Laboratory for Elementary-Particle Physics, Cornell University, Ithaca, NY 14853, USA}

\begin{abstract}
Recent results in studies of bottomonium especially relevant to
charmonium are reviewed.  This report covers dipion transition matrix
elements, $\eta$ production in $\Upsilon$ transitions, $\Upsilon$
decays to invisible particles, a search for a non-standard-model
pseudoscalar Higgs in $\Upsilon$ radiative decays, and $\Upsilon$
radiative decays to $\rm{f_2(1270)}$, $\eta$, and $\eta'$.
\end{abstract}

\maketitle

\thispagestyle{fancy}


\section{Introduction}
In describing bottomonium results at a charm conference, we implicitly
invoke heavy quark symmetry.  The QCD is supposed to be the same,
except that the bottom quark mass is about three times the charm quark
mass, the magnitude of the electric charge is half as big, and the
running strong coupling constant $\alpha_s$ is about 30\% smaller.
These differences will affect what portion of the effective potential
is explored, how well the non-relativistic approximation works, decay
multiplicities, and the number of bound quarkonium states, but the
changes should, in principle, be calculable.  This makes bottomonium a
different laboratory to study the same physics as seen in charmonium.
In this report, I will emphasize studies where bottomonium either
extends, checks, or suggests new studies that can be done in
charmonium.

\begin{figure*}
\centering
\includegraphics[width=135mm]{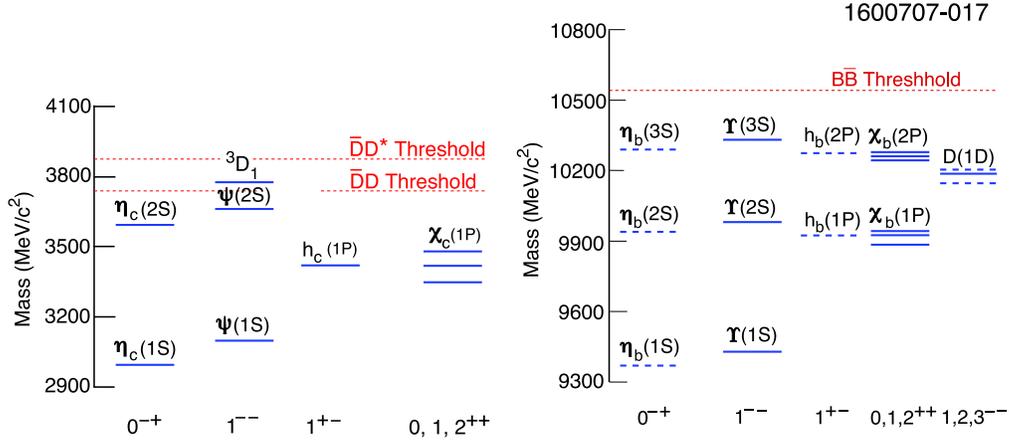}
\caption{The spectra of bound charmonium and bottomonium.  The vertical
scales are the same, offset to align the ground states.  States which have
not yet been observed are rendered as dashed lines.}
\label{oniumSpectra}
\end{figure*}

Figure~\ref{oniumSpectra} compares the bound spectra of charmonium and
bottomonium to illustrate some of these ideas.  You can see that the
bottomonium spectrum is richer, with more bound states and a wider
variety of decay scenarios.  It is also true that some of the
fundamental states, including the ground state, have not yet been
observed.

\section{Relevant Experiments}

The first $\Upsilon$ states were discovered in hadron collisions, and there
is still interesting work being done in bottomonium at the colliders.
In particular, D0 has recently measured polarization in
hadroproduction of bottomonium.  However, Jonas Rademacker discussed
this in detail in his report~\cite{Rademacker}, so I will not cover it here.

Direct production of bottomonia in $e^+ e^-$ collisions has been a
fruitful method for studying their properties.  CLEO has 6 million
$\Upsilon$(3S), 9 million $\Upsilon$(2S), and 21 million
$\Upsilon$(1S) events.  Belle collected 11 million $\Upsilon$(3S).  Of
course, both Belle and BaBar have hundreds of millions of
$\Upsilon$(4S) events.

With so much luminosity, Belle and BaBar also produce tens of millions
of $\Upsilon$(1S), $\Upsilon$(2S), and $\Upsilon$(3S) events with
initial state radiation, although these events are somewhat more difficult
to use effectively.

\section{Dipion Transition Matrix Element}

\begin{figure}
\centering
\includegraphics[width=80mm]{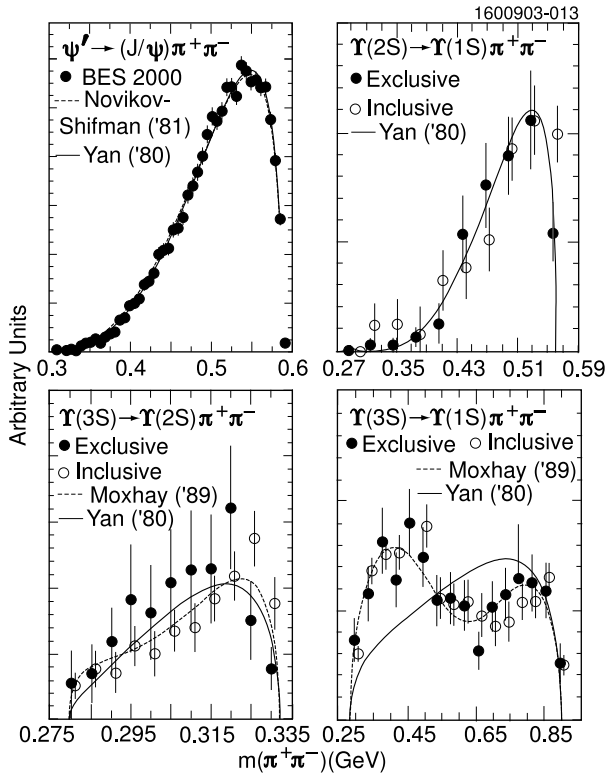}
\caption{Dipion mass distribution in
$\psi' \rightarrow\pi^+\pi^-J/\psi$ (upper left),
$\Upsilon(2S) \rightarrow \pi^+\pi^-\Upsilon(1S)$ (upper right),
$\Upsilon(3S) \rightarrow \pi^+\pi^-\Upsilon(2S)$ (lower left), and
$\Upsilon(3S) \rightarrow \pi^+\pi^-\Upsilon(1S)$ (lower right).
All distributions are well described by the Yan\cite{Yan} formulation,
which keeps only the $A$ term of Eq.~\ref{MatrixElement}, except the
last which has a distinctly different shape.  }
\label{oldDipion}
\end{figure}

\begin{figure}
\centering
\includegraphics[width=80mm]{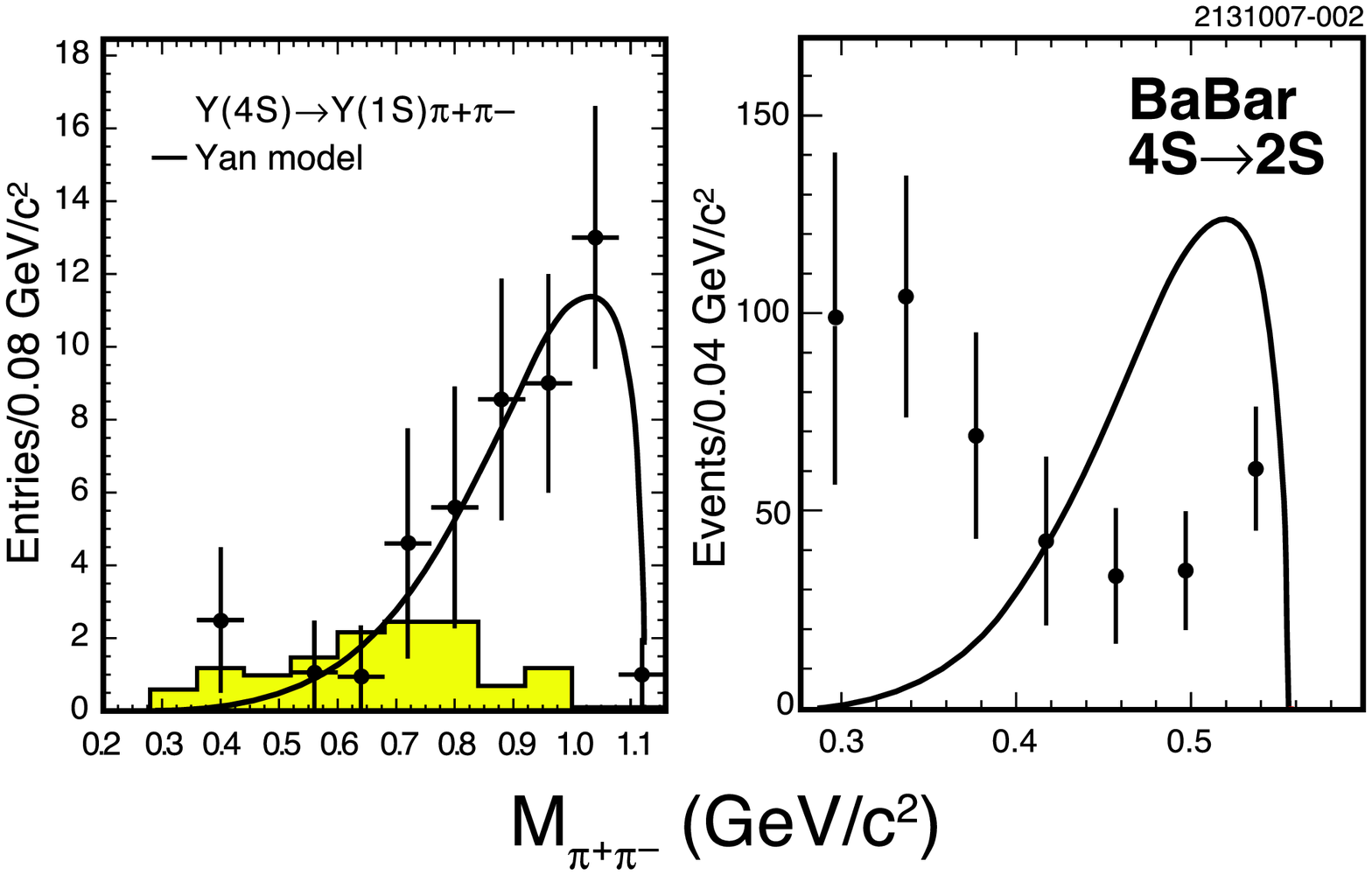}
\caption{Dipion mass distributions observed in $\Upsilon(4S)
\rightarrow \pi^+\pi^-\Upsilon(1S)$ by Belle~\cite{BelleDipion}
(left), and $\Upsilon(4S) \rightarrow \pi^+\pi^-\Upsilon(2S)$ by
BaBar~\cite{BaBarDipion} (right).  The former transition follows the
Yan formulation, but the latter is more complex.  The shaded histogram
in the Belle plot is an estimate of the background from $\Delta$M
sidebands.}
\label{4SDipion}
\end{figure}

For two decades, there has been a puzzle in the description of the
dipion mass distribution in bottomonium decays, as illustrated in
Fig.~\ref{oldDipion}.  While the dipion mass distribution for the
charmonium transition $\psi(2S) \rightarrow
\pi^+\pi^- J/\psi$, and the two bottomonium transitions
$\Upsilon(2S) \rightarrow \pi^+ \pi^- \Upsilon(1S)$ and  
$\Upsilon(3S) \rightarrow \pi^+ \pi^- \Upsilon(2S)$ are well described
by a single term in the matrix element~\cite{Yan}, the transition
$\Upsilon(3S) \rightarrow \pi^+ \pi^- \Upsilon(1S)$ shows a more
complicated, two hump structure.  More recently, data from
Belle~\cite{BelleDipion} and Babar~\cite{BaBarDipion} shown
in Fig.~\ref{4SDipion} add inputs to the
puzzle, showing that the decay
$\Upsilon(4S) \rightarrow \pi^+ \pi^- \Upsilon(1S)$ shows the simple
behavior, while
$\Upsilon(4S) \rightarrow \pi^+ \pi^- \Upsilon(2S)$ is more complex.

CLEO has recently attempted to approach the problem by analyzing
its $\Upsilon$ transition data in a two-dimensional Dalitz-like
fitting procedure~\cite{CLEODipion}.  Brown and Cahn~\cite{BrownCahn} describe
the relevant matrix element using PCAC and current algebra in the
general form
\begin{eqnarray}
\mathcal{M} & = & A(\epsilon'\cdot\epsilon)(q^2 - 2m_\pi^2) + \nonumber \\
&  &  B(\epsilon'\cdot\epsilon)E_1E_2 + \nonumber \\
&  & C[(\epsilon'\cdot q_1)(\epsilon\cdot q_2) + (\epsilon\cdot q_1)(\epsilon'\cdot q_2)],
\label{MatrixElement}
\end{eqnarray}
where $A, B,$ and $C$ are complex parameters, $\epsilon$ and
$\epsilon'$ are the $\Upsilon$ and $\Upsilon'$ polarization vectors,
$q^2=M_{\pi\pi}^2$, $E_1$ and $E_2$ are the pion energies, and $q_1$
and $q_2$ their four-momenta.

Instead of fitting in a single dimension $M_{\pi\pi}^2$, which is
equivalent to allowing only the $A$ term in the matrix element, CLEO
fits in $M_{\pi\pi}^2 - M_{\Upsilon\pi}^2$ space.  They assume that
the complex-valued $A$, $B$, and $C$ terms do not vary significantly
over the phase space of any of the decays.  In the fits they find that
the $C$ term is not needed, which is not unexpected because it
involves flipping the spin of the heavy $b$ quark. CLEO's limit is not
very stringent, $|C/A| < 1.09\%$ at 90\% confidence level, because the
shapes of the $C$ and $B$ terms are very similar.  If they set $C = 0$
they fit values for the ratio $B/A$ given in Table
{\ref{MatrixElements}}~\cite{CLEODipion}.

\begin{table}[ht]
\begin{center}
\caption{Fitted values for B/A in $\Upsilon$ dipion transitions}
\begin{tabular}{|c|c|c|c|}
\hline \textbf{Initial $\Upsilon$} & \textbf{Final $\Upsilon$} & \textbf{Re(B/A)} &
\textbf{Im(B/A)}
\\
\hline 3S & 1S & -2.52 $\pm$ 0.04 & $\pm$ 1.19 $\pm$ 0.06 \\
\hline 2S & 1S & -0.75 $\pm$ 0.15 & 0.00 $\pm$ 0.11 \\
\hline 3S & 2S & -0.40 $\pm$ 0.32 & 0.00 $\pm$ 1.10 \\
\hline
\end{tabular}
\label{MatrixElements}
\end{center}
\end{table}

Dubynskiy and Voloshin~\cite{Voloshin} argue that the CLEO
parametrization is too na{\"\i}ve because $B$ cannot possibly be
constant over the Dalitz plot.  So CLEO's fits are not universally
recognized as a solution to the dipion transition puzzle.  Yet they
show the power of the Dalitz technique, and it would be useful for
Belle in bottomonium and CLEO-c or BES in charmonium to see if this
technique describes the data well.

\section{Pseudoscalar Transitions}

In the charmonium system $\psi(2S) \rightarrow \eta J/\psi$ is
surprisingly large considering the small amount of available phase
space, with about a 3\% branching fraction~\cite{PDG}.  Yan~\cite{Yan}
predicted in 1980 that the rate in bottomonium should scale like
$\Gamma \propto (p^*)^3/m_Q^4$.  Kuang~\cite{Kuang} has more recently
refined this procedure and predicts\\ $B(\Upsilon(2S) \rightarrow \eta
\Upsilon(1S)) = (8.1 \pm 0.8)\times 10^{-4}$ and\\ $B(\Upsilon(3S)
\rightarrow \eta \Upsilon(1S)) = (6.7 \pm 0.7)\times 10^{-4}$.

CLEO has sought $\Upsilon(2S) \rightarrow \eta \Upsilon(1S)$ in the
final state where the
$\Upsilon$ decays $\Upsilon(1S) \rightarrow \mu\mu$ or $ee$ and
the $\eta$ decays $\eta \rightarrow \gamma\gamma$ or $\pi^+\pi^-\pi^0.$

\begin{figure}
\centering
\includegraphics[width=75mm]{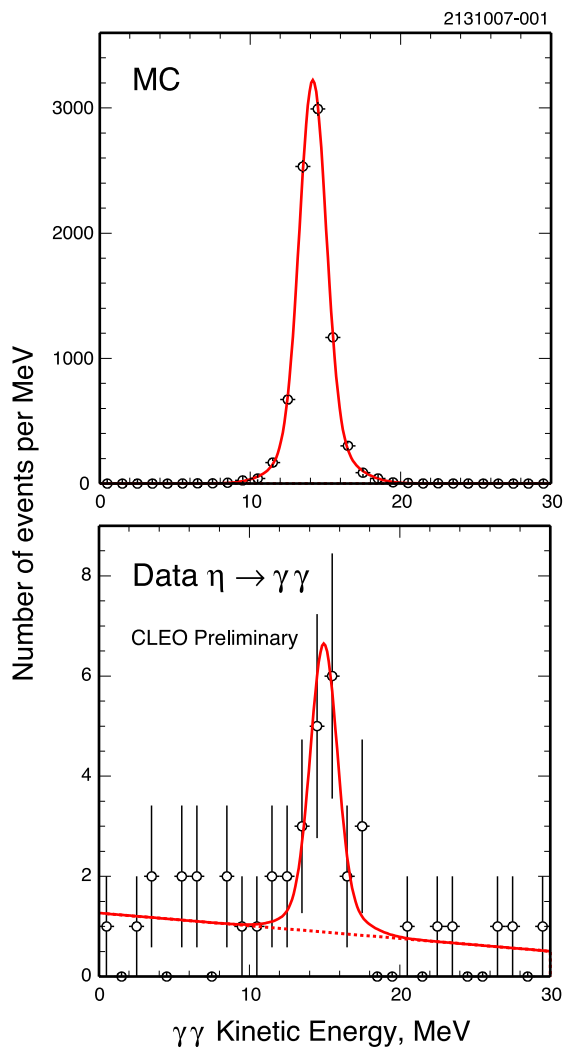}
\caption{The predicted and observed signal for $\eta$ production in $\Upsilon(2S) \rightarrow \eta \Upsilon(1S), \eta \rightarrow \gamma\gamma$.}
\label{2StoEta1S}
\end{figure}

CLEO plots the yield as a function of the kinetic energy of the $\eta$
candidate.  In this preliminary analysis CLEO sees a 5 standard
deviation peak in the kinetic energy of the $\gamma \gamma$ from
$\eta$ decay as shown in Fig.~\ref{2StoEta1S}.  This leads to a
(preliminary) branching fraction of \\$B(\Upsilon(2S) \rightarrow \eta
\Upsilon(1S)) = (2.5 \pm 0.7 \pm 0.5)\times 10^{-4}$,\\ somewhat
smaller but in the same general range as Kuang's prediction.  CLEO
also sees three events, with nearly zero expected background, in the
$\eta$ decay channel $\eta \rightarrow \pi^+\pi^-\pi^0$, which
corresponds to a consistent branching fraction.

Using the same technique, CLEO also seeks $\Upsilon(2S) \rightarrow
\pi^0 \Upsilon(1S)$, but sees no significant excess over background,
setting the (preliminary) 90\% confidence level upper limit
$B(\Upsilon(2S) \rightarrow \pi^0 \Upsilon(1S)) < 2.1 \times 10^{-4}.$
This is consistent with the expectation, obtained using Yan's simple
scaling prediction, that the $\pi^0$ rate should be 0.16 of the $\eta$
rate.

\section{$\Upsilon(1S)$ Decays to Invisible Particles}

The decays of charmonium or bottomonium states to undetectable
particles are a window on physics beyond the Standard Model.
McElrath~\cite{McElrath} has predicted that the neutralino $\chi$, a
dark matter candidate, could be produced in $\Upsilon(1S) \rightarrow
\chi\chi$ with a branching fraction of 0.41\%.  Possible production of
new gauge bosons or a light gravitino was described by
Fayet~\cite{Fayet}.  While it is true that $\Upsilon(1S) \rightarrow
\nu\nu$ via $Z^0$ production is a potential background, it is
calculated to be tiny enough as not to present a problem.

But how can you ``see'' these invisible decays?  The trick is to
produce a higher $\Upsilon$ state which decays via a two-pion
transition to the $\Upsilon(1S)$.  The experimentalist then uses the
two pions to both trigger the detector and to signal the presence of the
$\Upsilon(1S)$ with the missing mass recoiling against the dipion.
The remainder of the detector must be completely empty.

\begin{figure}
\centering
\includegraphics[width=80mm]{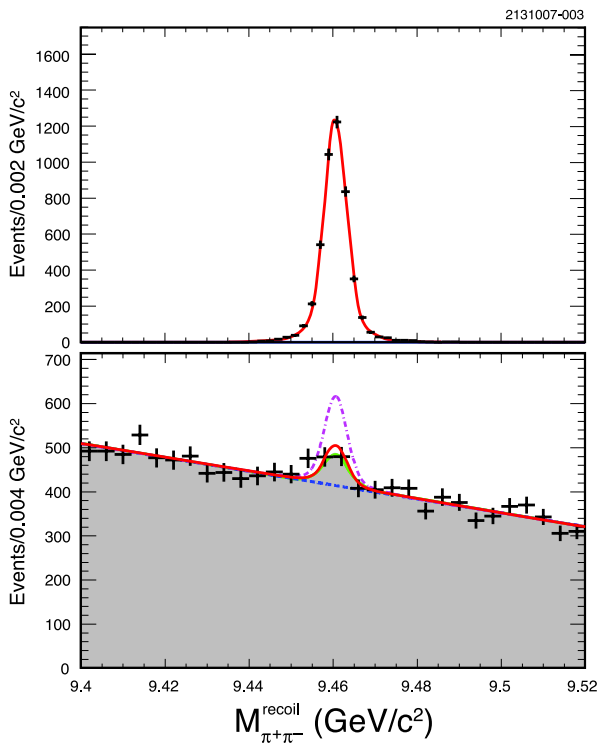}
\caption{The results of Belle's search for the decay sequence
$\Upsilon(3S) \rightarrow \pi^+\pi^-\Upsilon(1S), \Upsilon(1S)
\rightarrow$ ``invisible.''  The top panel shows the $\pi\pi$
invariant mass when $\Upsilon(1S) \rightarrow \mu\mu$, and the bottom
panel shows when the two pions and nothing more are seen in the
detector.  The peak seen in the lower plot is consistent with
expected backgrounds (grey).}
\label{BelleInvisible}
\centering
\end{figure}

\begin{figure}
\centering
\includegraphics[width=80mm]{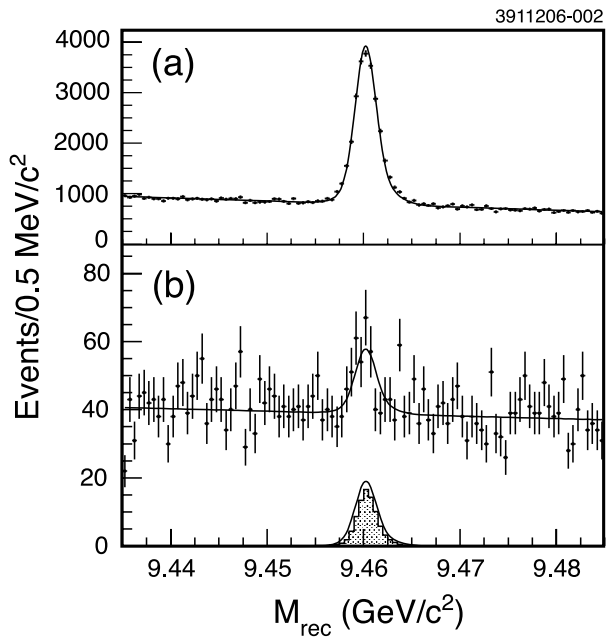}
\caption{The results of CLEO's search for the decay sequence
$\Upsilon(2S) \rightarrow \pi^+\pi^-\Upsilon(1S), \Upsilon(1S)
\rightarrow$ ``invisible.''  The top panel shows the $\pi\pi$
invariant mass in all $\Upsilon(2S)$ decays and the bottom panel shows
when the two pions and nothing more are seen in the detector.  The shaded
histogram at the bottom of the figure represents the expected background
and is consistent with what is observed.}
\label{CLEOInvisible}
\end{figure}

Figures \ref{BelleInvisible} and \ref{CLEOInvisible} show the results
of two searches.  Belle~\cite{BellePRL} uses $\Upsilon(3S)$ events so
the transition pions have enough energy to trigger the detector
reliably.  The top of Fig.~\ref{BelleInvisible} shows the dipion mass
spectrum when the $\Upsilon(1S) \rightarrow \mu\mu$ decay is observed,
to demonstrate the expected shape of a possible signal.  The bottom
part of the figure shows the dipion mass spectrum when the rest of the
detector shows no tracks and less than 3 GeV of neutral energy.

The CLEO data sample consists of 9 million $\Upsilon(2S)$ decays,
almost as large as Belle's 11 million $\Upsilon(3S)$ decays, and with
the advantage of a more favorable dipion branching rate to
$\Upsilon(1S)$.  Unfortunately, their two track trigger had to be
prescaled by a factor 20 to prevent saturating the data acquisition
system.  CLEO's results~\cite{CLEOInvis} are shown in
Fig.~\ref{CLEOInvisible}.  The top half shows the inclusive dipion
mass spectrum, and the bottom half shows what remains when the rest of
the detector is required to show no tracks and no photons of energy
more than 250 MeV.

A small peak is visible at the $\Upsilon(1S)$ mass in the bottom parts
of both figures.  In both cases, the observed peaking is consistent
with what is expected from Monte Carlo simulations where the decay
products of the $\Upsilon$ traveled down the beam line, thus escaping
the detector.  Both experiments thus set upper limits to the
production of invisible particles in $\Upsilon(1S)$ decays\\
$B(\Upsilon(1S) \rightarrow$ ``invisible'' $ < 0.25\%$ (Belle), and\\
$B(\Upsilon(1S) \rightarrow$ ``invisible'' $ < 0.39\%$ (CLEO).\\ Each
of these limits is an order of magnitude better than the previous best
limit.  Together, they set a limit on the branching fraction about
half the McElrath prediction for neutralino production, and better the
previous gravitino mass limit by a factor of four, to $m_{3/2} > 1.2
\times 10^{-7}$ eV.

Similar searches to these can be performed in the charmonium system,
where a much larger number of $\psi'$ events is available.  Of course,
the mass range that can be explored is more limited, and the predicted
branching fractions tend to be smaller, but such searches might be
fruitful for charmonium experiments to pursue.

\section{Radiative Decays of $\Upsilon(1S)$}
\subsection{Higgs Search}

In an effort to explain why the Higgs hasn't yet been seen, Dermisek,
Gunion, and McElrath~\cite{Dermisek} propose adding a
non-Standard-Model-like pseudoscalar Higgs $\rm{a_0}$ to the Minimal
Supersymmetric Standard Model (MSSM) to make it the ``Nearly MSSM''
(NMSSM).  This $\rm a_0$ must have mass less than twice the $b$ quark
mass, so that it can't decay to a pair of $b$ quark jets.  This proposal
explains the failure of the LEP experiments to see the Higgs at masses
up to 100 GeV, since the daughters of the Higgs decay can't make the b
jets those experiments sought.  Yet the hypothesis is natural, in the
sense that it avoids fine tuning of parameters to explain
observations.

The $\rm a_0$ should decay predominantly into $\tau\tau$ if it has
enough mass, and should be observable in $\Upsilon(1S) \rightarrow
\gamma \rm{a_0}$.

\begin{figure}
\centering
\includegraphics[width=80mm]{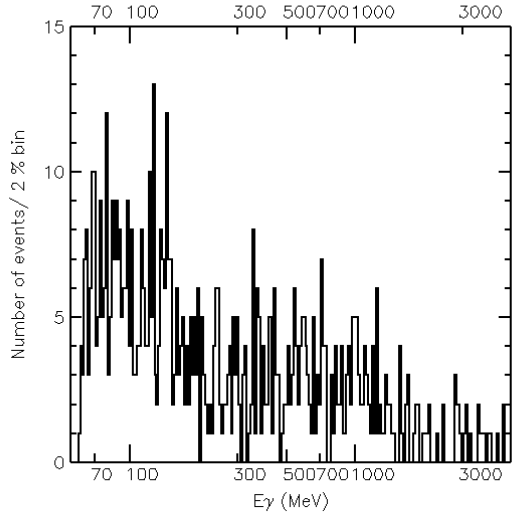}
\caption{The spectrum of photons in $\tau$-enriched $\Upsilon(1S)$
decays observed at CLEO.}
\label{PhotonSpectrum}
\end{figure}

\begin{figure}
\centering
\includegraphics[width=80mm]{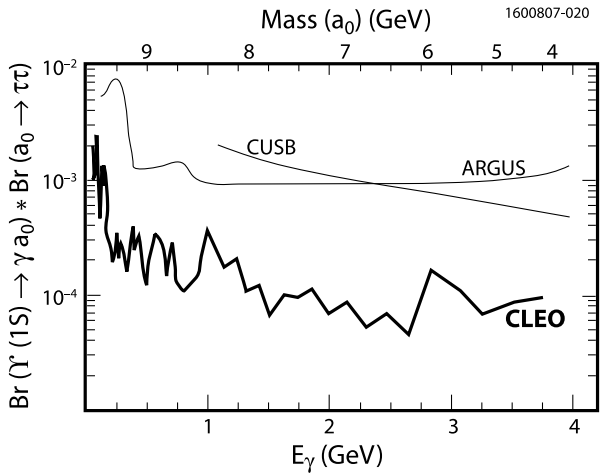}
\caption{Upper limits on the branching fraction of $\Upsilon(1S)
\rightarrow \gamma \rm{a_0}$ vs. photon energy (bottom scale)
and $\rm a_0$ mass (top scale).}
\label{HiggsUL}
\end{figure}

CLEO has sought these new Higgses by looking for monochromatic photons
in events likely to contain taus.  They tag $\Upsilon(1S)$ from
$\Upsilon(2S) \rightarrow \pi^+\pi^-\Upsilon(1S)$ to help eliminate
the copious QED backgrounds from $e^+ e^- \rightarrow \gamma\tau\tau$.
They flag the presence of $\tau$ pairs by seeking two 1-prong $\tau$
decays, one of which must be to a lepton, and by demanding missing
energy in the event.  The spectrum of photons they observe in such
events is shown in Fig.~\ref{PhotonSpectrum}, and leads to upper
limits shown in Fig.~\ref{HiggsUL}.  These upper limits improve on
older measurements by an order of magnitude or more, and rule out much
of the parameter space for NMSSM models.

\subsection{$\Upsilon(1S) \rightarrow \gamma \rm\bf{f_2(1270)}$}

\begin{figure}
\centering
\includegraphics[width=80mm]{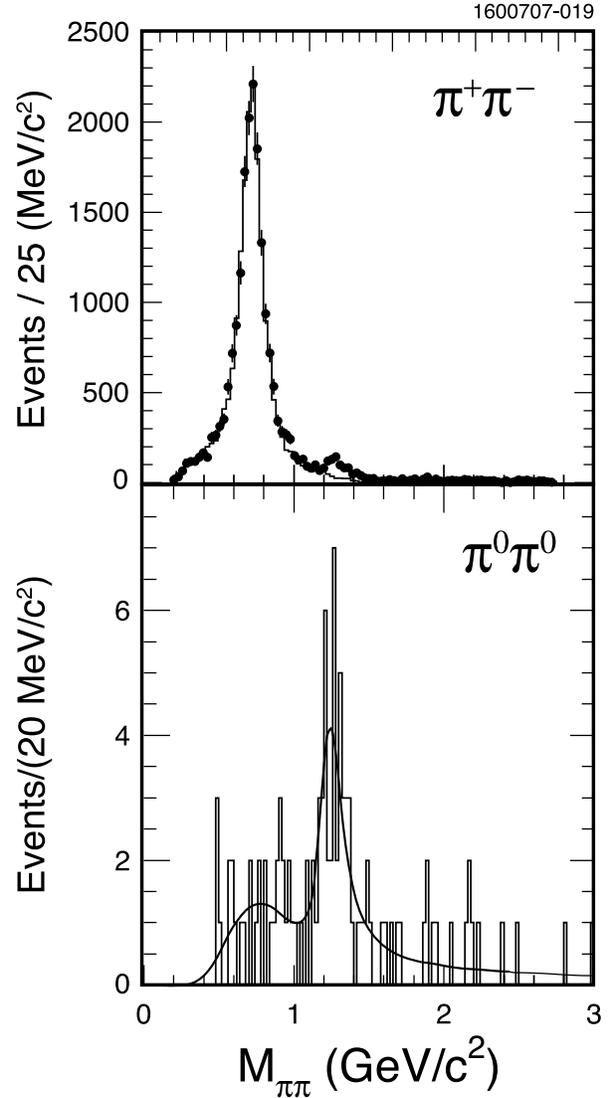}
\caption{CLEO observes $\Upsilon(1S) \rightarrow \gamma \rm{f_2(1270)}$
with the $\rm{f_2}$ decaying to charged pions (top) and neutral pions
(bottom).  The charged pion signal is contaminated with a huge
background from $e^+e^- \rightarrow \gamma\rho$, but the observed
small signal is confirmed in the neutral pion channel.}
\label{RadiativeF2}
\end{figure}

In the charmonium system, radiative decays are common and many
have been observed.  The decay $J/\psi \rightarrow \gamma \rm{f_2(1270)}$
is one of the most common.  In bottomonium, few exclusive radiative
decays are measured, but now CLEO has observed~\cite{CLEOf2}
$\Upsilon(1S) \rightarrow \gamma \rm{f_2(1270)}$ as can be seen in
Fig.~\ref{RadiativeF2}.

We test the simple heavy quark symmetry relation
\begin{eqnarray*}
\lefteqn{B(J/\psi \rightarrow \gamma \rm{f_2}) / B(\Upsilon(1S) \rightarrow \gamma \rm{f_2}) = \hspace{25mm}} \\
& \hspace{25mm}(q_c/q_b)^2(m_b/m_c)^2(\Gamma_{bb}/\Gamma_{cc}) \approx 20
\end{eqnarray*}
using the CLEO observations~\cite{CLEOf2}\\
$B(\Upsilon(1S) \rightarrow \gamma \rm{f_2}) =$
$(10.2 \pm 0.8 \pm 0.7) \times 10^{-5}$ ($\pi^+\pi^-$)\\
$B(\Upsilon(1S) \rightarrow \gamma \rm{f_2}) =
 (10.5 \pm 1.6 \pm 1.9) \times 10^{-5}$ ($\pi^0\pi^0$)\\
$B(\Upsilon(1S) \rightarrow \gamma \rm{f_2}) = 
 (10.23 \pm 0.97)\times 10^{-5}$ (combined).\\
The observed ratio \\
$B(J/\psi \rightarrow \gamma \rm{f_2}) / B(\Upsilon(1S) \rightarrow \gamma \rm{f_2}) = 14.0 \pm 1.7$\\
in satisfactory agreement with the expectations from scaling.

\subsection{$\Upsilon(1S) \rightarrow \gamma\eta'$ and $\gamma\eta$}

Does this success of scaling in radiative decay to $\rm{f_2}$ carry over
to other radiative decays?  Another prominent decay in the charmonium
system is
$B(J/\psi \rightarrow \gamma \eta' = (4.7 \pm 0.3) \times 10^{-3}$~\cite{PDG}.
Using the observed charm system decay rate ratio
$B(J/\psi \rightarrow \gamma \eta')/ B(J/\psi \rightarrow \gamma \rm{f_2}) = (3.4 \pm 0.4)$, and the relative rates of $\Upsilon$ and $J/\psi$ to $\rm{f_2}$, we
can predict the radiative decay rates for $\Upsilon(1S)$ to $\eta$ and $\eta'$.
The expectation is that these decays should be easily visible.

We already know that $\eta'$ is unconventional.  In radiative $J/\psi$
decay its branching fraction is five times as large as that for $\eta$.
There have been speculations that it might contain sizable gluon
content~\cite{KLOE}, or possible charmonium content.  Theorists have
used the vector dominance model (VDM)~\cite{Intemann}, $\eta_b$
mixing~\cite{Chao}, or higher twist to try to understand the unusual
behavior of the $\eta'$.

\begin{figure}
\centering
\includegraphics[width=80mm]{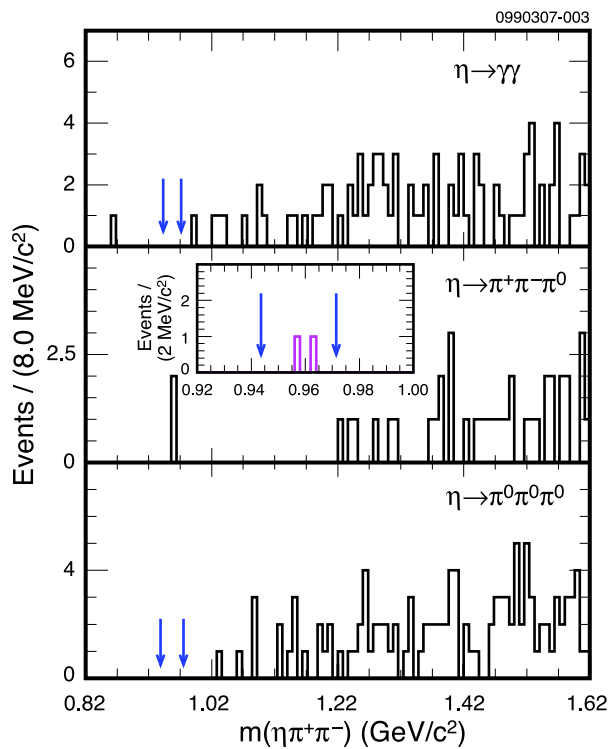}
\caption{CLEO seeks $\Upsilon(1S) \rightarrow \gamma \eta'$ with the
$\eta'$ decaying to $\pi^+\pi^-\eta$ and the daughter $\eta$ decaying
in any of three modes.  The blue arrows indicate where an expected $\eta'$
signal should be visible.  Two candidates are seen in the mode where
$\eta \rightarrow \pi^+\pi^-\pi^0$, but none are visible in the two
all-neutral $\eta$ decay modes, leading to the upper limit quoted in
the text.}
\label{RadiativeEta}
\end{figure}

CLEO has sought radiative decays to $\eta'$ and $\eta$ in 21 million
$\Upsilon(1S)$ decays with the $\eta'$ results shown in
Fig.~\ref{RadiativeEta}.  The upper limits they set are significantly
more stringent than former measurements,
$B(\Upsilon(1S) \rightarrow \gamma \eta') < 1.9 \times 10^{-6}$ and
$B(\Upsilon(1S) \rightarrow \gamma \eta) < 1.0 \times 10^{-6}$, whereas
na{\"\i}ve scaling as outlined above predicts $350 \times 10^{-6}$ and $70
\times 10^{-6}$, respectively.  So na{\"\i}ve scaling fails here by two
orders of magnitude.

These upper limits contradict the mixing approach of Chao~\cite{Chao}
by a factor of 30, but are still above the VDM predictions of
Intemann~\cite{Intemann} and the higher-twist description of
Ma~\cite{Ma}.

\section{Conclusion}

Bottomonium remains an active field of research at Fermilab, CLEO,
Belle and Babar.  I have presented new results in dipion transitions
among $\Upsilon$ states, $\eta$ and $\pi^0$ transitions in the
$\Upsilon$ system, searches for invisible particles and a new type of
Higgs, and radiative transitions to $\rm{f_2(1270)}$, $\eta$, and $\eta'$.
However, bottomonium studies are continuing, and more new results
can be expected next year.

\begin{acknowledgments}
Richard Galik was in the midst of preparing this talk when external
circumstances forced him to turn it over to me.  He had already chosen
most of the topics, prepared a few of the slides, and done much of the
research that went into the preparation of this note.  I want to thank
him for that work and the many helpful consultations we had.

This work was supported by the National Science
Foundation under contract NSF PHY-0202078. 

\end{acknowledgments}

\bigskip 

\end{document}